\documentclass[10pt]{iopart}

\usepackage{graphicx}
\usepackage{dcolumn}
\usepackage{bm}
\usepackage{color}

\begin{document}

\title[EIC in LHD]{Analysis of the MHD stability and energetic particles effects on EIC events in LHD plasma using a Landau-closure model}


\author{J. Varela}
\ead{jacobo.varela@nifs.ac.jp}
\address{National Institute for Fusion Science, National Institute of Natural Science, Toki, 509-5292, Japan}
\author{D. A. Spong}
\address{Oak Ridge National Laboratory, Oak Ridge, Tennessee 37831-8071, USA}
\author{L. Garcia}
\address{Universidad Carlos III de Madrid, 28911 Leganes, Madrid, Spain}
\author{S. Ohdachi}
\address{National Institute for Fusion Science, National Institute of Natural Science, Toki, 509-5292, Japan}
\author{K. Y. Watanabe}
\address{National Institute for Fusion Science, National Institute of Natural Science, Toki, 509-5292, Japan}
\author{R. Seki}
\address{SOKENDAI, Department of Fusion Science, Toki/Gifu and National Institute for Fusion Science, National Institute of Natural Science, Toki, 509-5292, Japan}

\date{\today}

\begin{abstract}
The aim of this study is to perform a theoretical analysis of the magnetohydrodynamic (MHD) stability and energetic particle effects on a LHD equilibria, calculated during a discharge where energetic-ion-driven resistive interchange mode (EIC) events were triggered. We use the reduced MHD equations to describe the linear evolution of the poloidal flux and the toroidal component of the vorticity in a full 3D system, coupled with equations of density and parallel velocity moments for the energetic particles species, including the effect of the acoustic modes, multiple energetic particles (EP) species, helical couplings and helically trapped EP. We add the Landau damping and resonant destabilization effects using a closure relation. The simulations suggest that the helically trapped EP driven by the perpendicular neutral beam injector (NBI) further destabilizes the $1/1$ MHD-like mode located at the plasma periphery ($r/a = 0.88$). If the $\beta$ of the EP driven by the perpendicular NBI is larger than $0.0025$ a $1/1$ EIC with a frequency around $3$ kHz is destabilized. If the effect of the passing EP driven by the tangential NBI is included on the model, any enhancement of the injection intensity of the tangential NBI below $\beta=0.025$ leads to a decrease of the instability growth rate. The simulations indicate that the perpendicular NBI EP is the main driver of the EIC events, as it was observed in the experiment. If the effect of the helical couplings are added in the model, an $11/13$ EIC is destabilized with a frequency around $9$ kHz, inward shifted ($r/a = 0.81$) compared to the $1/1$ EIC. Thus, one possible explanation for the EIC frequency chirping down from $9$ to $3$ kHz is a transition between the $11/13$ to the $1/1$ EIC due to a weakening of the destabilizing effect of the high $n$ modes, caused by a decrease of the EP drive due to a loss of helically trapped EP or a change in the EP distribution function after the EIC burst. The experimental data during the EIC bursting phase shows a complex mode structure and an inward shift of the instability, although no direct evidence of the proposed transition has been observed yet.

\end{abstract}

%
%
%
%
%


\vspace{2pc}

\maketitle

\ioptwocol

\section{Introduction \label{sec:introduction}}

The EIC events are observed in LHD discharges with low density and high ion temperature if the plasma is heated by strong perpendicular and tangential NBIs \cite{1}, showing magnetic fluctuations similar to the fishbone oscillations \cite{2,3}. In these discharges, the $\beta$ of the thermal plasma is small due to the low electron density and the large magnetic field strength, comparable with the $\beta$ of the EP driven by the perpendicular NBI. EIC events consist in a bursting instability with a frequency (f) around $9$ kHz triggered if the perpendicular NBI injection overcome some threshold, chirping down to $4$ kHz before stabilization \cite{4}.

Energetic particle driven instabilities can enhance the transport of fusion produced alpha particles, energetic hydrogen neutral beams and ion cyclotron resonance heated particles (ICRF) \cite{5,6,7}. The consequence is a decrease of the heating efficiency in helical devices such as LHD and W7-AS stellarators or tokamaks such as TFTR, JET and DIII-D \cite{8,9,10,11,12,13}. If the EP drift, bounce or transit frequencies are similar to the mode frequency a resonance takes place leading to an enhancement of the EP losses. In addition, plasma instabilities such as internal kinks \cite{14,15} or ballooning modes \cite{16,17} can be kinetically destabilized, as well as the energetic particle modes (EPM). these can be unstable for frequencies in the shear Alfven continua if the continuum damping is not strong enough to stabilize them \cite{18,19,20,21,22}. On the other hand, Alfv\' en Eigenmodes (AE) are driven in the spectral gaps of the shear Alfv\' en continua \cite{23,24}; these instabilities have been observed in several discharges and configurations \cite{25,26,27,28}. The different Alfv\' en eigenmode families ($n$ is the toroidal mode and $m$ the poloidal mode) are linked to frequency gaps produced by periodic variations of the Alfv\' en speed, for example: toroidicity induced Alfv\' en Eigenmodes (TAE) coupling $m$ with $m+1$ modes \cite{29,30,31,32}, beta induced Alfv\' en Eigenmodes driven by compressibility effects (BAE) \cite{33}, Reversed-shear Alfv\' en Eigenmodes (RSAE) due to local maxima/minima in the safety factor $q$ profile \cite{34,35}, Global Alfv\' en Eigenmodes (GAE) observed when there is an extremum in the Alfv\' en continua \cite{36,37,38}, ellipticity induced Alfv\' en Eigenmodes (EAE) coupling $m$ with $m+2$ modes \cite{39,40} and helical Alfv\' en Eigenmodes (HAE) where different toroidal modes are coupled \cite{41}.

Previous studies pointed out that the resistive interchange modes (RIC) are unstable in the magnetic hill region of LHD \cite{42,43,44,45,46}. The RIC can resonate with the precession motion of the helically trapped EP generated by the perpendicular NBI in the range of $f = 10$ kHz, leading to the enhancement of the EP radial transport \cite{47,48}. Such instability was identified as the trigger of the EIC events in previous studies \cite{2,4}. Consequently, the EIC events can be grouped in the family of the EPM.

LHD is a helical device heated by three NBI lines almost parallel to the magnetic axis with an energy of 180 keV and two NBI perpendicular to the magnetic axis with an energy of 32 keV. This study is limited to LHD discharges with Hydrogen plasma where the NBIs also inject Hydrogen; the EIC events are also observed in Deuterium plasma \cite{49}. The destabilization of EPM and AE caused by strong NBI injection was already observed in several LHD discharges \cite{19,32,34,38,41}.

The aim of the present study is to perform a theoretical analysis of the MHD stability and the destabilizing effects of the helically trapped EP on a LHD equilibria calculated during an EIC event. In the analysis we identify the threshold of the perpendicular NBI injection intensity to destabilize the EIC events, namely the $\beta$ of the helically trapped EP, adding in the simulations the effect of the acoustic modes, multiple energetic particles species and helical couplings. In addition, we analyze the structure of the instability eigenfunction to compare the simulation results with the experimental data.

A set of simulations are performed using an updated version of the FAR3D code \cite{50,51,52}, adding the moment equations of the energetic ion density and parallel velocity \cite{53,54}. This numerical model, with the appropriate Landau closure relations, solves the reduced non-linear resistive MHD equations including the linear wave-particle resonance effects, required for Landau damping/growth, and the parallel momentum response of the thermal plasma, required for coupling to the geodesic acoustic waves \cite{35}. The code follows the evolution of six field variables, starting from equilibria calculated by the VMEC code \cite{55}. A methodology has been developed to calibrate Landau-closure models against more complete kinetic models and optimize the closure coefficients \cite{35}. The model includes Landau resonance couplings, fast ion FLR \cite{54} and Landau damping of the modes on the background ions/electrons \cite{53}, although we did not consider the last two for simplicity.

This paper is organized as follows. The model equations, numerical scheme and equilibrium properties are described in section \ref{sec:model}. A study of the instability properties for different helically trapped EP configurations are presented in section \ref{sec:trapped}. A parametric analysis of the NBI injection intensity to identify the instability threshold is described in section \ref{sec:threshold}. Finally, the conclusions of this paper and the comparison with the experimental observations are presented in section \ref{sec:conclusions}.

\section{Equations and numerical scheme \label{sec:model}}

The reduced set of equations for high-aspect ratio configurations with moderate $\beta$-values (of the order of the inverse aspect ratio) is obtained following the method employed in Ref.\cite{56}, describing the evolution of the background plasma and fields retaining the toroidal angle variation. We obtain a reduced set of equations based upon an exact three-dimensional equilibrium that assumes closed nested flux surfaces. The moments of the fast ions kinetic equation truncated with a closure relation are added in the model to introduce the effect of the energetic particle population \cite{57}, describing the evolution of the energetic particle density ($n_{f}$) and velocity moments parallel to the magnetic field lines ($v_{||f}$). The coefficients of the closure relation are selected to match analytic TAE growth rates based upon a two-pole approximation of the plasma dispersion function.     

We assume high aspect ratio, medium $\beta$ (of the order of the inverse aspect ratio $\varepsilon=a/R_0$), small variation of the fields and small resistivity. The plasma velocity and perturbation of the magnetic field are defined as
\begin{equation}
 \mathbf{v} = \sqrt{g} R_0 \nabla \zeta \times \nabla \Phi, \quad\quad\quad  \mathbf{B} = R_0 \nabla \zeta \times \nabla \psi,
\end{equation}
where $\zeta$ is the toroidal angle, $\Phi$ is a stream function proportional to the electrostatic potential, and $\tilde \psi$ is the perturbation of the poloidal flux.

The equations, in dimensionless form, are
\begin{equation}
\frac{\partial \tilde \psi}{\partial t} =  \sqrt{g} B \nabla_\| \Phi  + \eta \varepsilon^2 J \tilde J^\zeta
\end{equation}
\begin{eqnarray} 
\frac{{\partial \tilde U}}{{\partial t}} =  -\epsilon v_{\zeta,eq} \frac{\partial U}{\partial \zeta} \nonumber\\
+ S^2 \left[{ \sqrt{g} B \nabla_\| J^\zeta - \frac{\beta_0}{2\varepsilon^2} \sqrt{g} \left( \nabla \sqrt{g} \times \nabla \tilde p \right)^\zeta }\right]   \nonumber\\
-  S^2 \left[{\frac{\beta_f}{2\varepsilon^2} \sqrt{g} \left( \nabla \sqrt{g} \times \nabla \tilde n_f \right)^\zeta }\right] 
\end{eqnarray} 
\begin{eqnarray}
\label{pressure}
\frac{\partial \tilde p}{\partial t} = -\epsilon v_{\zeta,eq} \frac{\partial p}{\partial \zeta} + \frac{dp_{eq}}{d\rho}\frac{1}{\rho}\frac{\partial \tilde \Phi}{\partial \theta} \nonumber\\
 +  \Gamma p_{eq}  \left[{ \sqrt{g} \left( \nabla \sqrt{g} \times \nabla \tilde \Phi \right)^\zeta - \nabla_\|  v_{\| th} }\right] 
\end{eqnarray} 
\begin{eqnarray}
\label{velthermal}
\frac{{\partial \tilde v_{\| th}}}{{\partial t}} = -\epsilon v_{\zeta,eq} \frac{\partial v_{||th}}{\partial \zeta} -  \frac{S^2 \beta_0}{n_{0,th}} \nabla_\| p 
\end{eqnarray}
\begin{eqnarray}
\label{nfast}
\frac{{\partial \tilde n_f}}{{\partial t}} = -\epsilon v_{\zeta,eq} \frac{\partial n_{f}}{\partial \zeta} - \frac{S  v_{th,f}^2}{\omega_{cy}}\ \Omega_d (\tilde n_f) - S  n_{f0} \nabla_\| v_{\| f}   \nonumber\\
- \varepsilon^2  n_{f0} \, \Omega_d (\tilde \Phi) + \varepsilon^2 n_{f0} \, \Omega_* (\tilde  \Phi) 
\end{eqnarray}
\begin{eqnarray}
\label{vfast}
\frac{{\partial \tilde v_{\| f}}}{{\partial t}} = -\epsilon v_{\zeta,eq} \frac{\partial v_{||f}}{\partial \zeta}  -  \frac{S  v_{th,f}^2}{\omega_{cy}} \, \Omega_d (\tilde v_{\| f}) \nonumber\\
- \left( \frac{\pi}{2} \right)^{1/2} S  v_{th,f} \left| \nabla_\|  v_{\| f}  \right| \nonumber\\
- \frac{S  v_{th,f}^2}{n_{f0}} \nabla_\| n_f + S \varepsilon^2  v_{th,f}^2 \, \Omega_* (\tilde \psi) 
\end{eqnarray}
Equation (2) is derived from Ohm’s law coupled with Faraday’s law, equation (3) is obtained from the toroidal component of the momentum balance equation after applying the operator $\nabla \wedge \sqrt{g}$, equation (4) is obtained from the thermal plasma continuity equation with compressibility effects and equation (5) is obtained from the parallel component of the momentum balance. Here, $U =  \sqrt g \left[{ \nabla  \times \left( {\rho _m \sqrt g {\bf{v}}} \right) }\right]^\zeta$ is the vorticity, $\rho_m$ the ion and electron mass density, $\rho = \sqrt{\phi_{N}}$ the effective radius with $\phi_{N}$ the normalized toroidal flux and $\theta$ the poloidal angle. The perturbation of the toroidal current density $\tilde J^{\zeta}$ is defined as:
\begin{eqnarray}
\tilde J^{\zeta} =  \frac{1}{\rho}\frac{\partial}{\partial \rho} \left(-\frac{g_{\rho\theta}}{\sqrt{g}}\frac{\partial \tilde \psi}{\partial \theta} + \rho \frac{g_{\theta\theta}}{\sqrt{g}}\frac{\partial \tilde \psi}{\partial \rho} \right) \nonumber\\
- \frac{1}{\rho} \frac{\partial}{\partial \theta} \left( \frac{g_{\rho\rho}}{\sqrt{g}}\frac{1}{\rho}\frac{\partial \tilde \psi}{\partial \theta} + \rho \frac{g_{\rho \theta}}{\sqrt{g}}\frac{\partial \tilde \psi}{\partial \rho} \right)
\end{eqnarray}
$v_{||th}$ is the parallel velocity of the thermal particles and $v_{\zeta,eq}$ is the equilibrium toroidal rotation. $n_{f}$ is normalized to the density at the magnetic axis $n_{f_{0}}$, $\Phi$ to $a^2B_{0}/\tau_{R}$ and $\Psi$ to $a^2B_{0}$. The radius $\rho$ is normalized to a generalized minor radius $a$; the resistivity to $\eta_0$ (its value at the magnetic axis); the time to the resistive time $\tau_R = a^2 \mu_0 / \eta_0$; the magnetic field to $B_0$ (the averaged value at the magnetic axis); and the pressure to its equilibrium value at the magnetic axis. The Lundquist number $S$ is the ratio of the resistive time to the Alfv\' en time $\tau_{A0} = R_0 (\mu_0 \rho_m)^{1/2} / B_0$. $\rlap{-} \iota$ is the rotational transform, $v_{th,f} = \sqrt{T_{f}/m_{f}}$ the energetic particle thermal velocity normalized to the Alfv\' en velocity in the magnetic axis $v_{A0}$ and $\omega_{cy}$ the energetic particle cyclotron frequency times $\tau_{A0}$. $q_{f}$ is the charge, $T_{f}$ the temperature and $m_{f}$ the mass of the energetic particles. The $\Omega$ operators are defined as:
\begin{eqnarray}
\label{eq:omedrift}
\Omega_d = \frac{\epsilon^2 \pi \rho^2 \omega_{b}}{d_{b}} \left[ \frac{\partial}{\partial \theta} \left( \frac{1}{\sqrt{g}} \right) \right]^{-1} \cdot \nonumber\\
\Bigg\{ \frac{1}{2 B^4 \sqrt{g}}  \left[  \left( \frac{I}{\rho} \frac{\partial B^2}{\partial \zeta} - J \frac{1}{\rho} \frac{\partial B^2}{\partial \theta} \right) \frac{\partial}{\partial \rho}\right] \nonumber\\
-   \frac{1}{2 B^4 \sqrt{g}} \left[ \left( \rho \beta_* \frac{\partial B^2}{\partial \zeta} - J \frac{\partial B^2}{\partial \rho} \right) \frac{1}{\rho} \frac{\partial}{\partial \theta} \right] \nonumber\\ 
+ \frac{1}{2 B^4 \sqrt{g}} \left[ \left( \rho \beta_* \frac{1}{\rho} \frac{\partial B^2}{\partial \theta} -  \frac{I}{\rho} \frac{\partial B^2}{\partial \rho} \right) \frac{\partial}{\partial \zeta} \right] \Bigg\}
\end{eqnarray}

\begin{eqnarray}
\label{eq:omestar}
\Omega_* = \frac{1}{B^2 \sqrt{g}} \frac{1}{n_{f0}} \frac{d n_{f0}}{d \rho} \left( \frac{I}{\rho} \frac{\partial}{\partial \zeta} - J \frac{1}{\rho} \frac{\partial}{\partial \theta} \right) 
\end{eqnarray}
Here the $\Omega_{d}$ operator is the average drift velocity of a helically trapped particle and $\Omega_{*}$ models the diamagnetic drift frequency. The parameter $\omega_{b}$ indicates the bounce frequency and $d_{b}$ the bounce length of the helically trapped EP guiding center. For more details regarding the derivation of the average drift velocity operator please see the Appendix. It should be noted that the effect of the helical particle trapping is only considered through a modification of the average drift velocity, and the simulations are still based on an isotropic model. Future efforts will be oriented to improve the present model including anisotropic distributions, leading to a more accurate analysis.

We also define the parallel gradient and curvature operators as
\begin{equation}
\label{eq:gradpar}
\nabla_\| f = \frac{1}{B \sqrt{g}} \left( \frac{\partial \tilde f}{\partial \zeta} +  \rlap{-} \iota \frac{\partial \tilde f}{\partial \theta} - \frac{\partial f_{eq}}{\partial \rho}  \frac{1}{\rho} \frac{\partial \tilde \psi}{\partial \theta} + \frac{1}{\rho} \frac{\partial f_{eq}}{\partial \theta} \frac{\partial \tilde \psi}{\partial \rho} \right)
\end{equation}
\begin{equation}
\label{eq:curv}
\sqrt{g} \left( \nabla \sqrt{g} \times \nabla \tilde f \right)^\zeta = \frac{\partial \sqrt{g} }{\partial \rho}  \frac{1}{\rho} \frac{\partial \tilde f}{\partial \theta} - \frac{1}{\rho} \frac{\partial \sqrt{g} }{\partial \theta} \frac{\partial \tilde f}{\partial \rho}
\end{equation}
with the Jacobian of the transformation,
\begin{equation}
\label{eq:Jac}
\frac{1}{\sqrt{g}} = \frac{B^2}{\varepsilon^2 (J+ \rlap{-} \iota I)}
\end{equation}

Equations~\ref{pressure} and~\ref{velthermal} introduce the parallel momentum response of the thermal plasma. These are required for coupling to the geodesic acoustic waves, accounting for the geodesic compressibility in the frequency range of the geodesic acoustic mode (GAM) \cite{58,59}.

Equilibrium flux coordinates $(\rho, \theta, \zeta)$ are used. Here, $\rho$ is a generalized radial coordinate proportional to the square root of the toroidal flux function, and normalized to the unity at the edge. The flux coordinates used in the code are those described by Boozer \cite{60}, and $\sqrt g$ is the Jacobian of the coordinate transformation. All functions have equilibrium and perturbation components represented as: $ A = A_{eq} + \tilde{A} $.

The FAR3D code uses finite differences in the radial direction and Fourier expansions in the two angular variables. The numerical scheme is semi-implicit in the linear terms.

The present model was already used to study the AE activity in LHD \cite{31,61}, TJ-II \cite{62,63,64} and DIII-D \cite{65,66,67}, indicating reasonable agreement with the observations.

\subsection{Equilibrium properties}

We use fixed boundary results from the VMEC equilibrium code \cite{55} calculated using the LHD reconstruction of a discharge during an EIC event (shot 116190). The electron density and temperature profiles were reconstructed by Thomson scattering data and electron cyclotron emission. Table~\ref{Table:1} shows the main parameters of the thermal plasma and table~\ref{Table:2} the details of the EP driven by the perpendicular and tangential NBIs. Both NBIs inject hydrogen. Thus, the cyclotron frequency is the same, $\omega_{cy} = 2.41 \cdot 10^{8}$ s.

\begin{table}[t]
\centering
\begin{tabular}{c c c c}
\hline
$T_{i}$ (keV) & $n_{i}$ ($10^{20}$ m$^{-3}$) & $\beta_{th}$ & $V_{A}$ ($10^{6}$ m/s) \\ \hline
3.41 & 0.72 & 0.002 & 6.44 \\
\end{tabular}
\caption{Thermal plasma properties (values at the magnetic axis). The first column is the thermal ion temperature, the second column is the thermal ion density, the third column is the thermal $\beta$ and the fourth column is the Alfv\' en velocity.} \label{Table:1}
\end{table}

\begin{table}[t]
\centering
\begin{tabular}{c}
Perpendicular NBI EP \\ 
\end{tabular}

\begin{tabular}{c c}
\hline
$n_{f,\perp}$ ($10^{20}$ m$^{-3}$) & $\beta_{f,\perp}$\\ \hline
0.0035 & 0.0013 \\ 
\end{tabular}

\begin{tabular}{c}
Parallel NBI EP \\ 
\end{tabular}

\begin{tabular}{c c}
\hline
$n_{f,||}$ ($10^{20}$ m$^{-3}$) & $\beta_{f,||}$\\ \hline
0.0039 & 0.0025 \\
\end{tabular}
\caption{Properties of the EP driven by the perpendicular (first row) and tangential (second row) NBIs (values at the magnetic axis). First column is the energetic particle density and the second column is the energetic particle $\beta$.} \label{Table:2}
\end{table}

\begin{figure}[h!]
\centering
\includegraphics[width=0.45\textwidth]{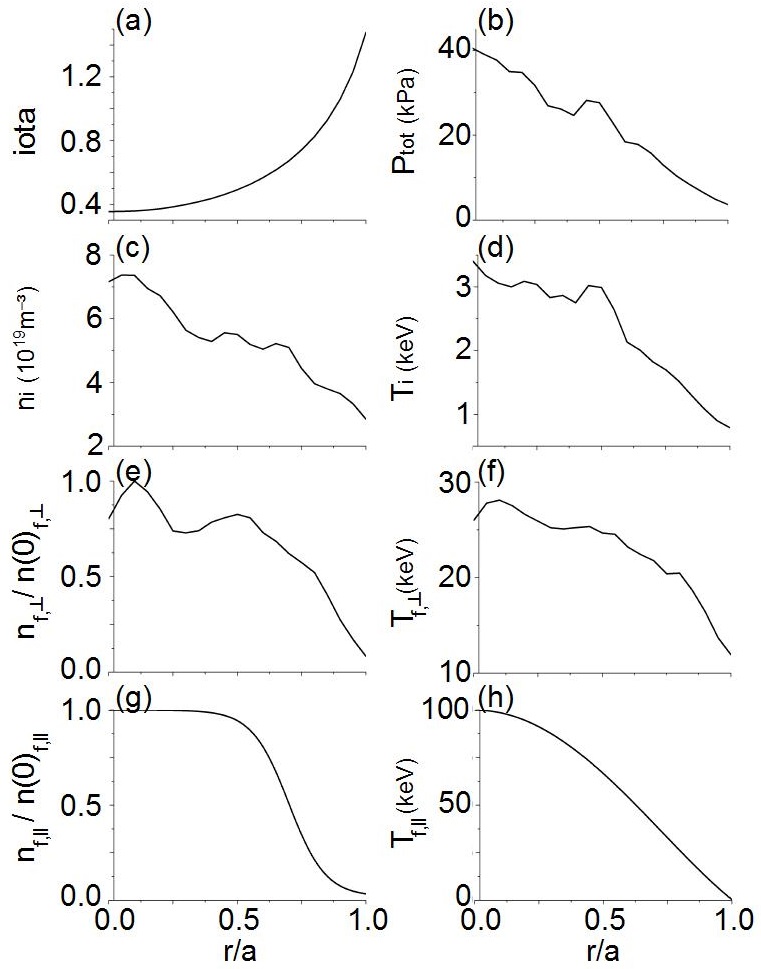}
\caption{(a) Iota profile, (b) equilibrium pressure (thermal + EP pressure), (c) thermal plasma density, (d) thermal plasma temperature, (e) perpendicular NBI EP density, (f) perpendicular NBI EP temperature, (g) parallel NBI EP density and (h) parallel NBI EP temperature.}\label{FIG:1}
\end{figure}

The magnetic field at the magnetic axis is $2.5$ T and the averaged inverse aspect ratio is $\varepsilon=0.16$. The energy of the injected particles by the perpendicular NBI is $T_{f,\perp}(0) = 40$ keV, but we take the nominal energy $T_{f,\perp}(0) = 28$ keV ($v_{th,f} = 1.64 \cdot 10^{6}$ m/s) resulting in an averaged Maxwellian energy equal to the average energy of a slowing-down distribution. In the same way, the energy of the injected particles by the tangential NBI is $T_{f,||}(0) = 180$ keV but we take the nominal energy $T_{f,||}(0) = 100$ keV ($v_{th,f} = 3.1 \cdot 10^{6}$ m/s). Figure~\ref{FIG:1}  (a) shows the iota profile, (b) the equilibrium pressure (thermal plasmas + EP pressure), (c) the thermal plasma density, (d) the thermal plasma temperature, (e) the perpendicular NBI EP density, (f) the perpendicular NBI EP temperature, (g) the parallel NBI EP density and (h) the parallel NBI EP temperature. It should be noted that the effect of the equilibrium toroidal rotation is not included in the model for simplicity. The effect of the Doppler shift on the instability frequency caused by the toroidal rotation is small, particularly for a mode located in the plasma periphery, as it is observed in the experiments.

\begin{figure}[h!]
\centering
\includegraphics[width=0.4\textwidth]{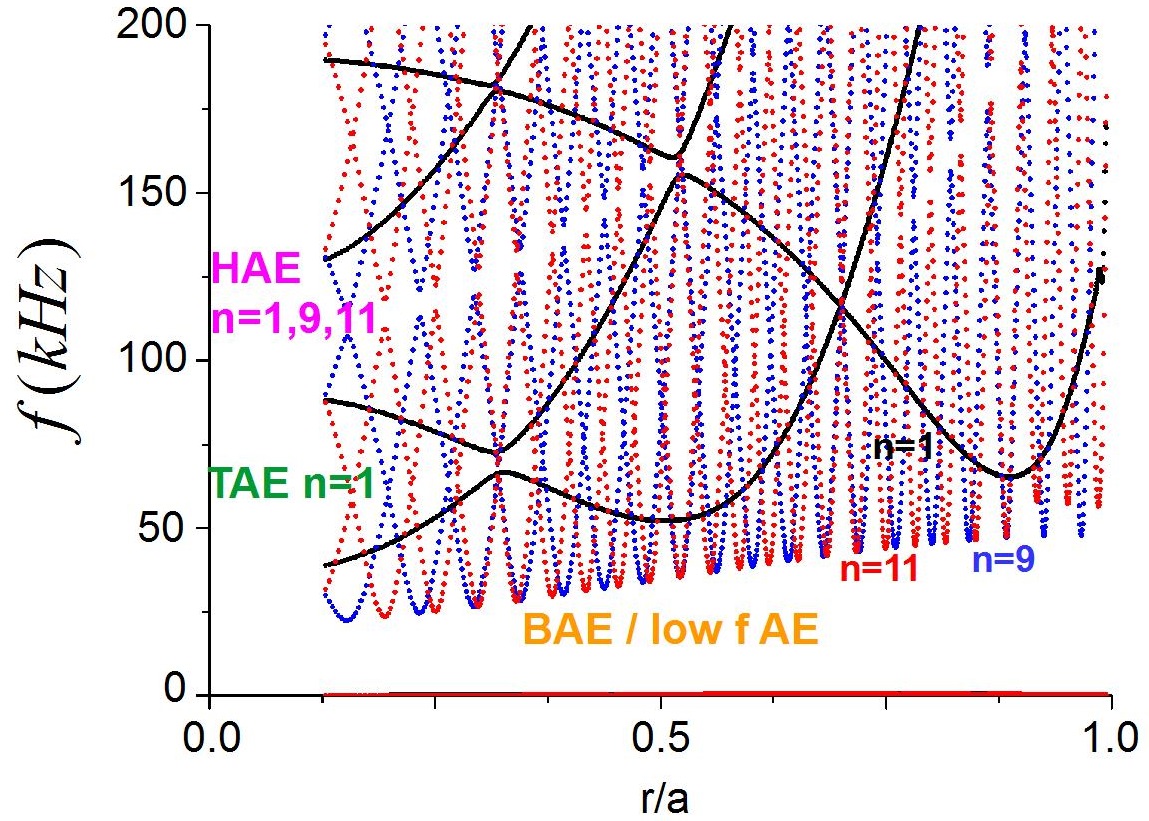}
\caption{Alfv\' en gaps of the $n=1,9,11$ helical family. The black dots indicate the $n=1$ modes, the blue dots show the $n=9$ modes and the red dots show the $n=11$ modes.}\label{FIG:2}
\end{figure}

Figure~\ref{FIG:2} shows the Alfv\' en gaps for the helical coupled family $n=1,9,11$ (b). There are three gaps: the BAE or low frequency AE gap below $40$ kHz that covers all the normalized minor radius, the $n=1$ TAE gap in the range of $f = [65,75]$ kHz at the inner plasma region as well as the $n=1,9,11$ HAE gap in the range of $f = [110,180]$ kHz between the inner-middle plasma region. The analysis of the continuum gaps includes the effect of the sound waves. 

\subsection{Simulations parameters}

The dynamic and equilibrium toroidal (n) and poloidal (m) modes included in the study are summarized in table~\ref{Table:3} for the simulations with toroidal or helical couplings. The simulations are performed with a uniform radial grid of 1000 points. The plasma core is not included in the analysis ($\rho < 0.3$) to avoid the destabilization of undesired modes near the magnetic axis, because the aim of the study is to analyze modes destabilized in the plasma periphery. In the following, the mode number notation is $n/m$, which is consistent with the $\iota=n/m$ definition for the associated resonance.

\begin{table}[h]
\centering
\begin{tabular}{c}
Simulation with toroidal couplings \\
\end{tabular}

\begin{tabular}{c || c | c}
\hline
n & 1 & 0  \\ \hline
m & $[1,2]$ & $[0,6]$ \\ \hline
\end{tabular}

\begin{tabular}{c}
Simulation with helical couplings \\
\end{tabular}

\begin{tabular}{c || c c c | c c}
\hline
n & 1 & 9 & 11 & 0 & 10 \\ \hline
m & $[1,8]$ & $[6,18]$ & $[6,22]$ & $[0,4]$ & $[-7,3]$ \\ \hline
\end{tabular}
\caption{Dynamic and equilibrium toroidal (n) and poloidal (m) modes in the simulations with toroidal couplings (upper row) and in the simulations with helical coupling (lower row).} \label{Table:3}
\end{table}

The usual MHD parities are broken by the closure of the kinetic moment equations (6) and (7). Consequently, both parities must be included in the model $sin(m\theta + n\zeta)$ and $cos(m\theta + n\zeta)$ for all dynamic variables, and allowing for both a growth rate and a real frequency in the eigenmode time series analysis. The convention of the code is, in the case of the pressure eigenfunction, that $n > 0$ corresponds to the Fourier component $\cos(m\theta + n\zeta)$ and $n < 0$ to $\sin(-m\theta - n\zeta)$. For example, the Fourier component for mode $-1/2$ is $\cos(-1\theta + 2\zeta)$ and for the mode $1/-2$ is $\sin(-1\theta + 2\zeta)$. The magnetic Lundquist number is $S=5\cdot 10^6$ which similar to the experimental value in the middle plasma.

\begin{figure}[h!]
\centering
\includegraphics[width=0.45\textwidth]{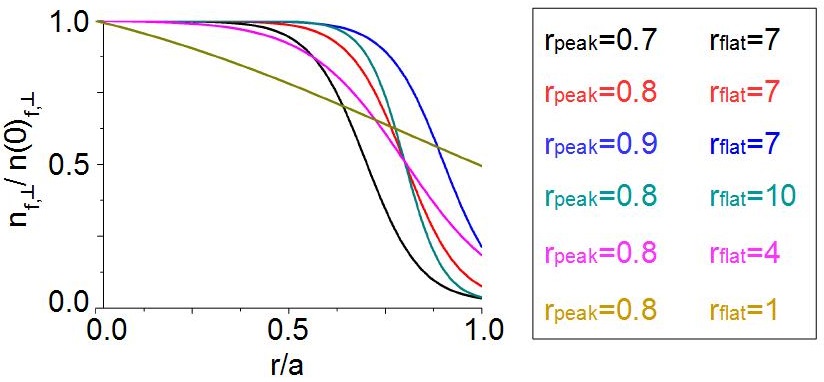}
\caption{Example of the analytic density profiles of the perpendicular NBI EP used in the parametric study.}\label{FIG:3}
\end{figure}

The $\beta_{f}$ value introduces the density ratio between the EP and bulk plasma ($n_{f}(0)/n_{e}(0)$) at the magnetic axis into the model. The EP density and temperature of the perpendicular beam are calculated by the code MORH \cite{68,69}. For the tangential NBI we use the same parameters as in reference \cite{61}. In addition, we study the effect of the location of the EP density profile gradient (NBI deposition region) and profile flatness using analytic expressions for the density profile of the perpendicular NBI EP. The analytic expression is the following:

\begin{equation}
\label{EP_dens}
$$n_{f,||}(r) = \frac{(0.5 (1+ \tanh(r_{flat} \cdot (r_{peak}-r))+0.02)}{(0.5 (1+\tanh(r_{flat} \cdot r_{peak}))+0.02)}$$
\end{equation}
The location of the gradient is controlled by the parameter ($r_{peak}$) and the flatness by ($r_{flat}$). Figure~\ref{FIG:3} shows some examples of the perpendicular NBI EP density profiles used in the study.

The resonance coupling efficiency between the EIC and the energetic particles is determined by the ratio between the EP thermal velocity and the Alfv\' en velocity at the magnetic axis ($v_{th,f}/v_{A0}$). The EP thermal velocity is associated with the beam energy (NBI voltage). The Landau closure model used here is based on two moment equations for the fast ions, which is equivalent to a two-pole approximation to the plasma dispersion relation. The closure coefficients are adjusted by fitting analytic AE growth rates. This translates to a Lorentzian energy distribution function; to lowest order the Lorentzian can be matched either to a Maxwellian or a slowing-down distribution by choosing an equivalent average energy. For the results given in this paper, we have matched the EP temperature to the mean energy of a slowing-down distribution function. A more precise matching to the resonance function for a slowing-down distribution can be obtained by including higher moment equations for the fast ions. FAR3D has recently been extended to three and four moment versions, for improved matching to a variety of non-Maxellian distributions. These require further testing and calibration, using the methods presented in \cite{35}. They will be the topic of future research.

\section{Instability resonance versus helically trapped EP configuration \label{sec:trapped}}

\begin{figure}[h!]
\centering
\includegraphics[width=0.5\textwidth]{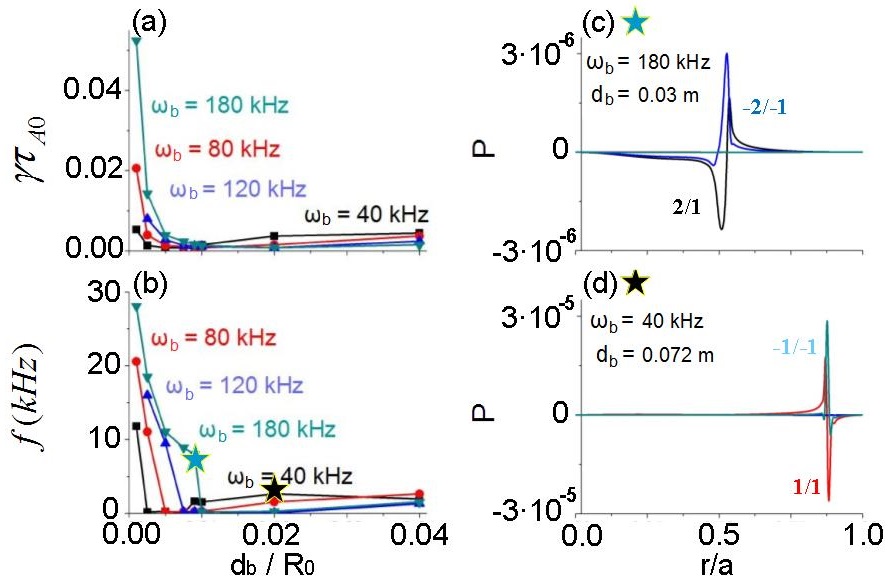}
\caption{Instability growth rate (a) and frequency (b) driven by helically trapped EP with different guiding center bounce frequencies and lengths. The black and blue stars indicate instabilities driven by deeply helically trapped EP (c) and helically trapped EP (d).}\label{FIG:4}
\end{figure}

We analyze in this section the effect of the guiding center bounce frequency and length on the instability properties. Figure~\ref{FIG:4} shows the instability growth rate, frequency and eigenfunction structure for different bounce frequency and length values (fixed $\beta_{f,\perp} = 0.015$).

The analysis shows two different resonances. On one side, there is a resonance driven by deeply helically trapped EP ($d_{b}/R_{0} < 0.01$) that takes place if the $v_{b} > 2.3 \cdot 10^{3} m/s$, leading to the destabilization of the $1/2$ mode in the middle plasma and an instability frequency of $f = [8,28]$ kHz. On the other hand, there is another broad resonance if $d_{b}/R_{0} > 0.01$ and the bounce frequency is $\omega < 80$ kHz, leading to the destabilization of the $1/1$ mode in the plasma periphery and an instability frequency around $4$ kHz ($v_{b} < 2.3 \cdot 10^{3} m/s$). The growth rate of the instability driven by the deeply trapped EP is up to 10 times higher. In the following, we only consider the resonance where the $1/1$ mode is destabilized because the experimental data indicates that this is the mode involved in the EIC events. In addition, following the guiding center of the helically trapped particle orbits using the 3D Monte Carlo code GCR \cite{70}, we found that the bounce length of the helically trapped EP guiding center is several centimeters and the bounce frequency is $\omega \le 100$ kHz, so $v_{b} \approx 2 \cdot 10^3$, consistent with the resonance regime of helically trapped EP, not deeply helically trapped EP. The analysis of the resonance driven by the deeply trapped EP will be the topic of a future study. 

\begin{figure}[h!]
\centering
\includegraphics[width=0.5\textwidth]{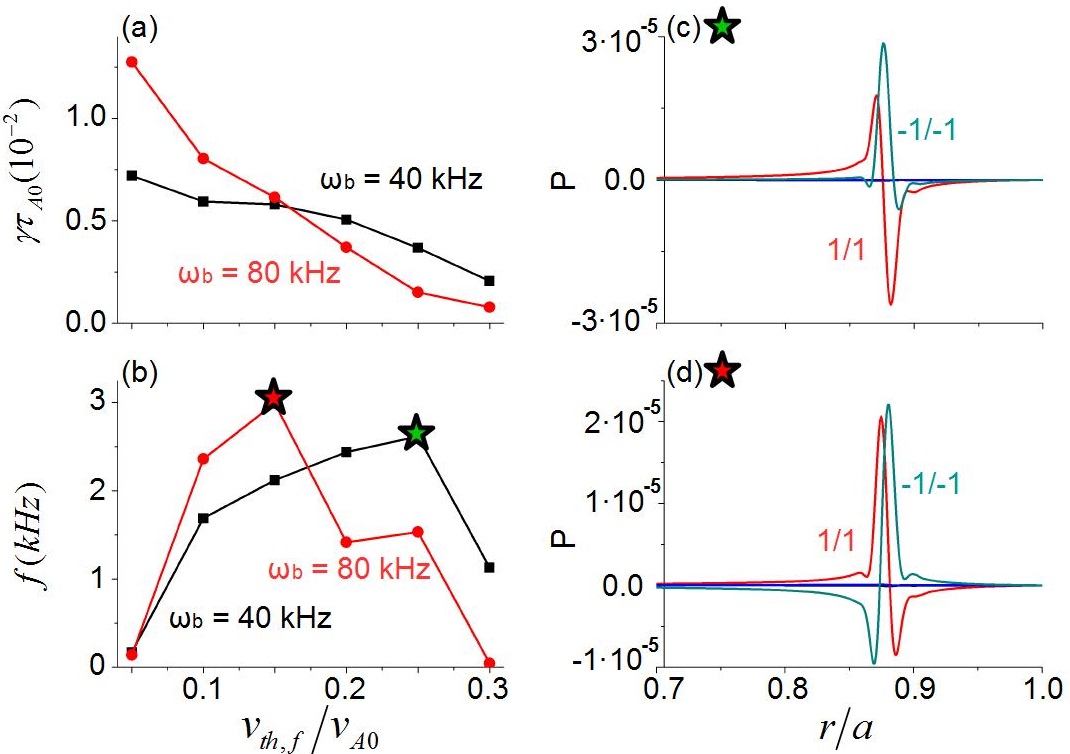}
\caption{Instability growth rate (a) and frequency (b) driven by helically trapped EP with different $v_{th,f}/v_{A0}$ ratios. The green and red stars indicate the instabilities driven if the bounce frequency is (c) $40$ kHz and (d) $80$ kHz.}\label{FIG:5}
\end{figure}

The next step of the study consists of analyzing the resonance efficiency of the EP to destabilize the instability given the energy of the EP (the ratio $v_{th,f}/v_{A0}$). Figure~\ref{FIG:5} shows the instability growth rate, frequency and eigenfunction structure for different $v_{th,f}/v_{A0}$ ratios.

The instability growth rate increases as the $v_{th,f}/v_{A0}$ ratio decreases, although the highest frequency is observed for $v_{th,f}/v_{A0} = 0.15$ if $\omega_{b} = 80$ kHz and $v_{th,f}/v_{A0} = 0.25$ if $\omega_{b} = 40$ kHz. In both cases the mode $1/1$ is destabilized in the plasma periphery showing a clear asymmetry between parities, pointing out that the instability is caused by the destabilizing effect of the EP. The experimental data indicates that the EP involved in the EIC events do not have time to undergo a significant slowing down process, thus they are weakly thermalized when the resonance takes place. Thus, the simulations with a bounce frequency of $\omega_{b} = 40$ kHz are expected to be closer to experimental conditions. Because the highest frequency takes place at $v_{th,f}/v_{A0} = 0.25$, this may characterize the EP with a smaller slowing down time compared to the $\omega_{b} = 80$ kHz case. In the following, we select the helically trapped EP configuration with $\omega_{b} = 40$ kHz, $d_{b} = 0.072$ m and $v_{th,f}/v_{A0} = 0.25$.

\begin{figure}[h!]
\centering
\includegraphics[width=0.5\textwidth]{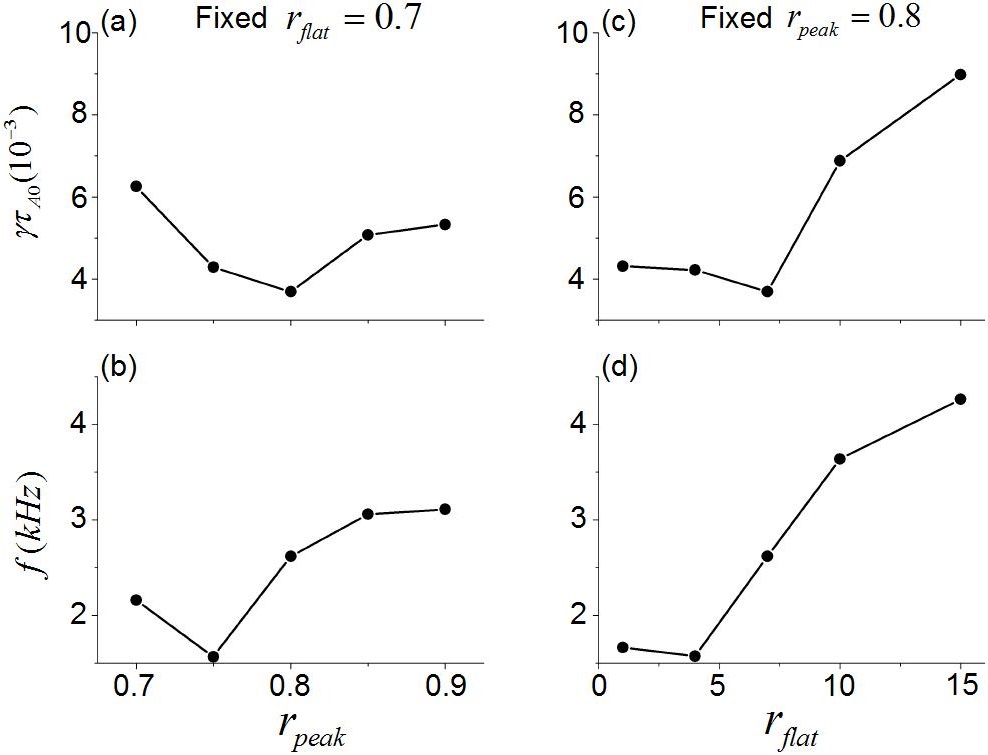}
\caption{Instability growth rate (a) and frequency (b) for different locations of the EP density gradient $r_{peak}$. (c) Instability growth rate and (d) frequency for different flatness of the EP density profile $r_{flat}$.}\label{FIG:6}
\end{figure}

Next, we study the effect of the EP density profile on the instability properties. Figure~\ref{FIG:6} shows the instability growth rate and frequency for several analytic density profiles defined by equation~\ref{EP_dens} (see figure~\ref{FIG:3}).

The instability growth rate increases if the EP density gradient is displaced inward or outward from $r_{peak} = 0.8$, and the instability frequency is larger if the EP density gradient is located at $r_{peak} > 0.8$. The perpendicular NBI deposition region is located in the outer plasma region thus the model configuration that better represents the experiments are the simulations with $r_{peak} > 0.8$. On the other hand, the instability growth rate and frequency decrease if the gradient of the EP density profiles is weaker because the source of free energy to destabilize EP driven modes is linked to the gradient of the EP density profile. Consequently, in the following we use an analytic expression of the density profile with $r_{peak} = 0.85$ and $r_{flat} = 10$ to maximize the destabilizing effect of the EP while keeping an EP density profile representative of the experiment.

\section{Study of the instability threshold \label{sec:threshold}}

\begin{figure*}[h!]
\centering
\includegraphics[width=0.8\textwidth]{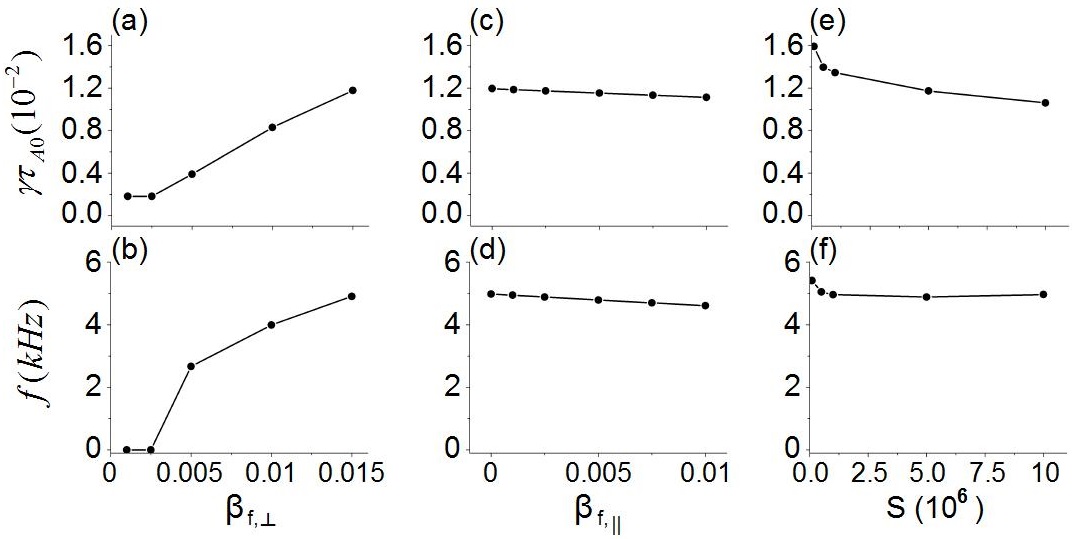}
\caption{Instability growth rate (a) and frequency (b) for different values of $\beta_{f,\perp}$ (fixed $\beta_{f,||} = 0.002$ and $S = 5 \cdot 10^{6}$). Instability growth rate (c) and frequency (d) for different values of $\beta_{f,||}$ (fixed $\beta_{f,\perp} = 0.015$ and $S = 5 \cdot 10^{6}$). Instability growth rate (e) and frequency (f) for different values of the thermal plasma resistivity (fixed $\beta_{f,||} = 0.002$ and $\beta_{f,\perp} = 0.015$).}\label{FIG:7}
\end{figure*}

In this section we analyze the effect of the injection intensity of the perpendicular and parallel NBI, identifying the $\beta$ threshold of the helically trapped EP driven by the perpendicular NBI ($\beta_{f,\perp}$) and the passing EP driven by the tangential NBI ($\beta_{f,||}$) to destabilize AE. In order to perform this study, we include in the numerical model two new equations that introduce the perturbation caused by the passing EP. The new set of equations have the same structure as the equations \ref{nfast} and \ref{vfast} although the correction of the averaged drift velocity operator is not included and the averaged energies are different. Figure~\ref{FIG:7} shows the instability growth rate and frequency for different $\beta_{f,\perp}$ and $\beta_{f,||}$ values as well as different thermal plasma resistivities. It should be noted that the simulations are performed including both EP populations in order to reproduce the combined effects of the perpendicular and tangential NBI EP on the instability threshold.

\begin{figure*}[h!]
\centering
\includegraphics[width=0.9\textwidth]{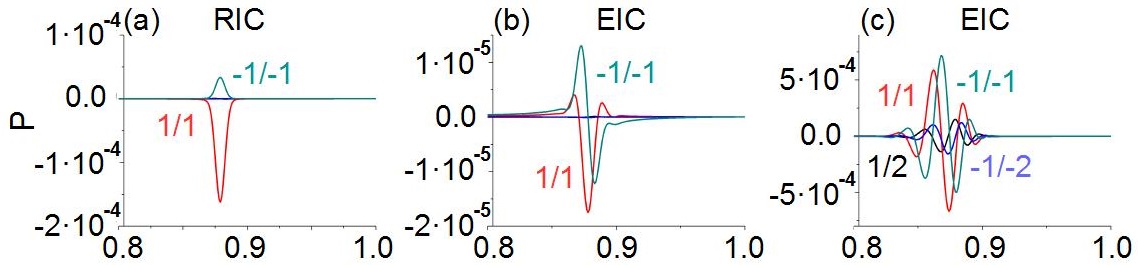}
\caption{Instability eigenfunction if (a) $\beta_{f,\perp} = 0.0025$, (b) $\beta_{f,\perp} = 0.005$ and $S = 5 \cdot 10^6$, (c) $\beta_{f,\perp} = 0.005$ and $S = 10^5$.}\label{FIG:8}
\end{figure*}

For a fixed injection intensity of the tangential beam to $\beta_{f,||} = 0.002$, if $\beta_{f,\perp} > 0.0025$ there is a large increase of the instability growth rate and frequency, Figure~\ref{FIG:7}a and b, pointing out a threshold in $\beta_{f,\perp}$ linked to a transition between different instabilities. Below this threshold a $1/1$ MHD mode at $r/a = 0.88$ is unstable, identified by its low frequency (below $1$ kHz) and the eigenfunction structure where both mode parities are similar (asymmetries in the mode parities are caused by the destabilizing effect of the EP). See figure~\ref{FIG:8}a. The $1/1$ MHD mode calculated in the simulation is similar to the precursor MHD activity of the EIC events \cite{4}. Above the threshold, the $1/1$ EIC at $r/a = 0.88$ is destabilized with a $f > 3$ kHz showing asymmetries in the mode parities. See figure~\ref{FIG:8}b. This instability is similar to the perturbation observed in the experiment before and after the bursting phase of the EIC events. 

Now we analyze the effect of the tangential NBI EP with the injection intensity of the perpendicular beam fixed at $\beta_{f,\perp} = 0.015$, Figure~\ref{FIG:7}c and d. Increasing $\beta_{f,||}$ leads to a decrease of the instability growth rate and frequency. The $\beta_{f,||}$ threshold to drive an instability is higher than $0.01$, the highest value used in this study, although the characteristic of the AE destabilized by the tangential NBI EP can be analyzed increasing the $\beta_{f,||}$ up to $0.025$, a drive strong enough to destabilize a $1/2$ BAE in the middle plasma region with a frequency of $23$ kHz. The decrease of the $1/1$ EIC growth rate is caused by the different resonance properties of the passing EP with the thermal plasma, leading to a stabilizing effect over the perturbation driven by the helically trapped EP that we refer to as the multiple beam damping effect. Such a damping effect was already observed in TFTR experiments where the NBI driven EP provided damping that prevented the destabilization of Alpha particle driven AEs \cite{71,72,73} except in the afterglow phase of the discharge. It should be noted that the simulations are coherent with the experimental observations that identified an increase of the time-wise spacing of the EIC events as the injection intensity of the tangential NBI increases. In addition, the simulations also indicate that the $1/1$ EIC is destabilized if $\beta_{f,||} = 0$, in line with the experimental observations, because the EIC events are only triggered if the perpendicular NBI injection intensity is large enough to overcome the EIC threshold, even if the tangential NBI injection is weak, but never in discharges with plasma heated only by tangential beams.

To confirm the destabilization of the EIC above the $\beta_{f,\perp}$ threshold we analyze the dependence of the instability growth rate and frequency on the thermal plasma resistivity. See Figure~\ref{FIG:7}e and f. A large plasma resistivity must lead to an enhancement of the instability growth rate if the mode is MHD-like. On the other hand, if the plasma resistivity is small enough the mode is stabilized. If the simulation resistivity is reduced by a factor of two, increasing the magnetic Lundquist number up to $S = 10^{7}$, the mode is not stabilized and its growth rate decreases less than $10 \%$ keeping almost the same instability frequency. On the other hand, increasing the simulation resistivity by a factor of 50 ($S = 10^5$) leads to an increase of the instability growth rate by around $30 \%$, although the instability frequency increases less than $5 \%$. Consequently, even if the instability shows some dependency on the thermal plasma resistivity, it is weaker than an MHD-like mode. Figure~\ref{FIG:8}c shows the instability eigenfunction structure for the simulation with the largest plasma resistivity ($S = 10^{5}$), indicating an increase of the eigenfunctions width and the destabilization of the $1/2$ modes. Hence the perturbation is enhanced, although the eigenfunction structure is similar compared to the simulations with lower resistivity, pointing out that the same type of instability is involved in both cases. See figure~\ref{FIG:8}b.

\subsection{Effect of the helical couplings}

\begin{figure}[h!]
\centering
\includegraphics[width=0.45\textwidth]{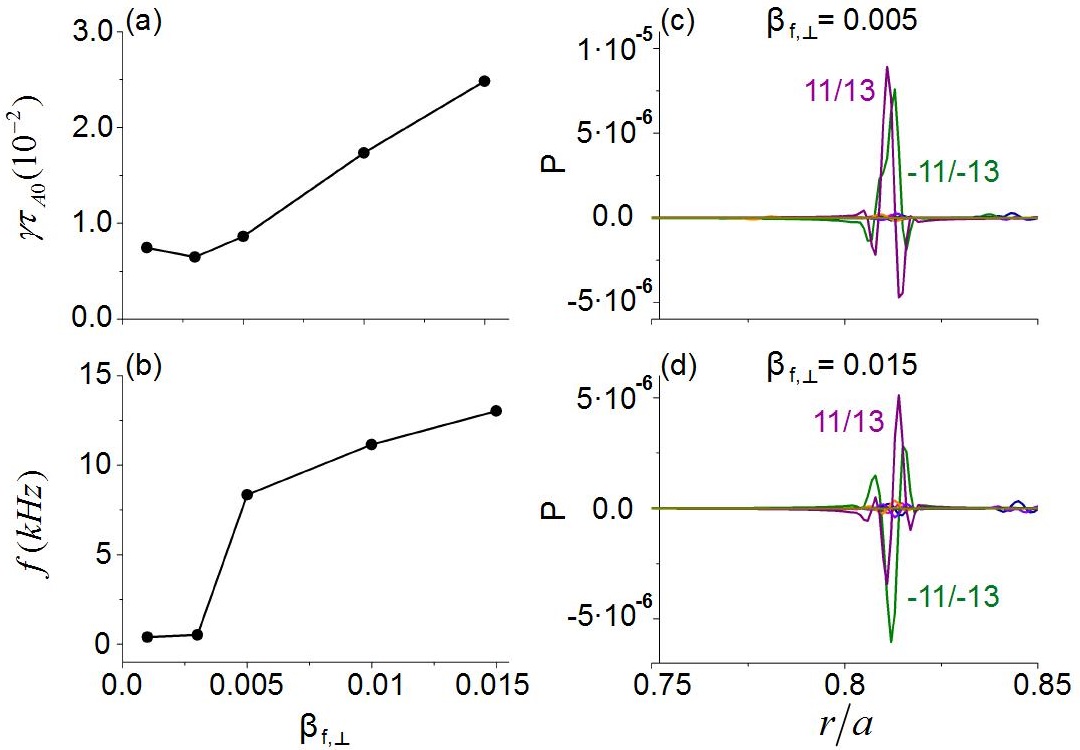}
\caption{Instability growth rate (a) and frequency (b) for different values of $\beta_{f,\perp}$ including the effect of the helical couplings (fixed $\beta_{f,||} = 0.002$). Instability eigenfunction if (c) $\beta_{f,\perp} = 0.005$ and (d) $\beta_{f,\perp} = 0.015$.}\label{FIG:9}
\end{figure}

Next, we study the effect of the helical couplings on the instability growth rate, frequency and eigenfunction structure. See figure~\ref{FIG:9}.

\begin{figure}[h!]
\centering
\includegraphics[width=0.45\textwidth]{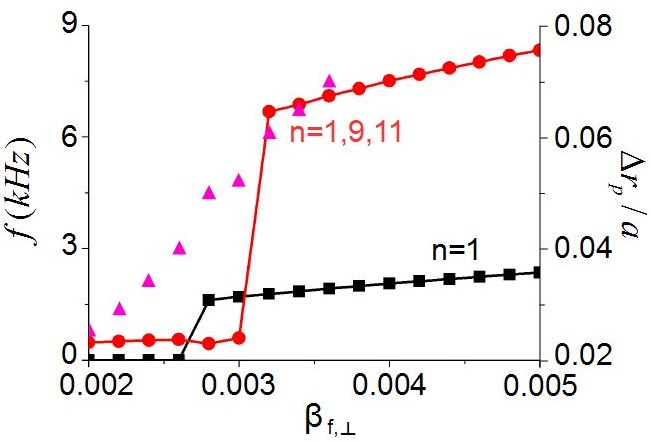}
\caption{Instability frequency for different values of $\beta_{f,\perp}$ near the transition between MHD-like and EIC for the $n=1$ toroidal family (black line) and helical family $n=1,9,11$ (red line) fixed $\beta_{f,||} = 0.002$. The pink triangles show the instability eigen-function width (normalized to the minor radius).}\label{FIG:10}
\end{figure}

Again, if $\beta_{f,\perp} = 0.005$ (fixed $\beta_{f,||} = 0.002$) an $11/13$ EIC is destabilized, displaced inward ($r/a = 0.82$) and also showing a more complex structure compared to the simulations with only toroidal couplings. See figure~\ref{FIG:9}c. In addition, the instability frequency is larger, around the $8.5$ kHz. Any further increment of $\beta_{f,\perp}$ leads to a larger growth rate and frequency, although the eigenfunction structure of the $11/13$ EIC is similar. See figure~\ref{FIG:9}d. It should be noted that if the simulation is performed below the $\beta_{f,\perp}$ threshold to destabilize the $11/13$ EIC, a $11/13$ MHD like mode is unstable. Figure~\ref{FIG:10} shows the instability frequency driven by the $n=1$ toroidal family (black line) and helical family $n=1,9,11$ (red line), identifying with increased accuracy the $\beta_{f,\perp}$ value required to trigger the transition between the MHD-like modes and EIC. The transition for the $n=1$ toroidal family takes place if $\beta_{f,\perp} = 0.0028$ although for the helical family $n=1,9,11$ is observed if $\beta_{f,\perp} = 0.0032$, pointing out that the $1/1$ EIC is destabilized by a lower EP driving regarding $11/13$ EIC. Consequently, the EIC is first destabilized by the $1/1$ mode and a further increase of the $\beta_{f,\perp}$ leads to the destabilization of the high $n$ modes of the $n=1,9,11$ helical family triggering an $11/13$ EIC. The normalized width of the instability eigen-function (pink triangles) increases with $\beta_{f,\perp}$. Below the $\beta_{f,\perp}$ threshold to destabilize the $1/1$ EIC ($\beta_{f,\perp}=0.0026$), the normalized eigen-function width of the $1/1$ RIC is $\Delta r_{p} / a = 0.04$, similar to the instability width measured in the experiment during the phase I and early phase II. Above the $1/1$ EIC $\beta_{f,\perp}$ threshold the eigen-function width keeps increasing, up to $\Delta r_{p} / a= 0.07$ for $\beta_{f,\perp}=0.0036$, consistent with the inward extension of the instability observed in the experiment during the late phase II and the transition to the phase III.

The $11/13$ EIC calculated by the simulation shows some analogies with the observations during the bursting phase of the EIC event, for example, the instability frequency range, the inward shifted location and the complex mode structure. In addition, the simulations suggest a transition from the $1/1$ EIC to the $11/13$ EIC caused by an increment of the EP driving, although there is no experimental evidence that such transition takes place. Dedicated experiments are required to confirm the proposed scenario by the numerical modeling.

\section{Conclusions and discussion \label{sec:conclusions}}

The set of linear simulations performed by the FAR3d code identified several instabilities whose features are compatible with the experimental observations during different phases of the EIC events.

The FAR3d model is updated to include the effect of the helically trapped energetic particles in the averaged drift velocity operator. The new formulation of the averaged drift velocity operator is used to study the destabilizing effect of the EP driven by the perpendicular NBI on LHD plasma, including characteristics of the motion of the helically trapped particle, in particular the bounce length and frequency of the guiding center.

Two different resonances are identified regarding the bounce length and frequency of the helically trapped particle guiding center. One resonance is caused by deeply helically trapped EP with a fast bounce frequency ($\omega_{b} \ge 120$ kHz), leading to the destabilization of a $1/2$ EIC or BAE in the central plasma region with frequencies between $f = [8,28]$ kHz. The other resonance is associated with helically trapped EP with a bounce frequency $\omega_{b} \le 80$ kHz, leading to the destabilization of a $1/1$ EIC at $r/a = 0.88$ with a frequency around $f = 4$ kHz. The instabilities caused by the helically trapped resonance are further analyzed due to the similarities with perturbations observed during the EIC events.

The simulations show an unstable $1/1$ MHD-like mode at $r/a = 0.88$ similar to the precursor MHD activity observed before the triggering of the EIC events, identified as a resistive interchange mode. An increase of the perpendicular NBI injection intensity leads to a further destabilization of the $1/1$ MHD-like mode until a threshold at $\beta_{f,\perp} = 0.0028$ is overcome, leading to the triggering of a $1/1$ EIC with a frequency around $3$ kHz, also located at $r/a = 0.88$. The transition from the $1/1$ MHD-like mode to the $1/1$ EIC is consistent with phase I of the EIC event as it was defined in reference~\cite{4}, as well as the threshold of the perpendicular NBI injection intensity observed in the experiment. As the EP drive is increased (larger $\beta_{f,\perp}$) the $1/1$ EIC is further destabilization and the perturbation extends inwards, similar to the phase II of the EIC event. If the EP driving is further increased up to $\beta_{f,\perp} = 0.0032$ and the effect of the helical couplings is added in the simulations, an $11/13$ EIC is destabilized at $r/a = 0.82$ with a frequency around $f=8.5$ kHz, showing a complex eigenfunction structure. Such an instability has several features in common with the perturbation observed during the bursting phase III of the EIC events. It should be noted that the mode structure in the experiment is very complex and present analysis is limited to linear simulations where only the dominant mode is considered. Thus, the model only reproduces part of the complexity of the process, identifying the different EIC phases with the destabilization of a $1/1$ RIC (phase I), a $1/1$ EIC (phase II), an $11/13$ EIC (phase III) and again a $1/1$ EIC (phase IV), as a simplification of the experimental observations. Consequently, to assess the problem in a more comprehensive way non linear simulations that include the effect of the subdominant modes are required. Nevertheless, the present study is already able to evaluate the major modes involved in the EIC event. In addition, the simulations without helical couplings assume that the helical couplings are weak because the high $n$ modes of the $n=1,9,11$ helical family are EIC stable, thus the analysis of the $n=1$ toroidal family represents in first approximation the EIC phenomena before and after the EIC bursting phase.

The simulations also suggest an explanation for the frequency chirp down observed in the experiment, showing a transition from the $11/13$ to the $1/1$ EIC. Such a transition takes place because the EP driving effect decreases due to a loss of helically trapped particles \cite{47,48}, lower $\beta_{f,\perp}$, or the modification of the helical trapped particle distribution function after each EIC burst. A more detailed analysis of the effect of the helically trapped distribution function during the EIC events will be the topic of a future study. It should be noted that there is no experimental evidence of the transition proposed by the numerical modeling, although the simulations reproduce several features of the EIC bursting phase. This scenario must be confirmed on dedicated experiments.

The simulations with multiple EP populations, particularly the helically trapped energetic particles driven by the perpendicular NBI and the passing energetic particles driven by the tangential NBI, show a stabilizing effect of the tangential beam as the injection intensity increases, reducing the growth rate of the EIC destabilized by the perpendicular beam. A similar behavior was observed in the experiments because the EIC events show a larger time-wise spacing as the injection intensity of the tangential NBI increases, pointing out a stabilizing effect over the perpendicular NBI driving. The multiple beam damping effect is caused by the different resonance properties of the passing energetic particles with the thermal plasma compared to the helically trapped energetic particles. The threshold to trigger an AE driven by the tangential NBI is $\beta_{f,||} = 0.025$, higher than that which characterized perpendicular NBI. If this threshold is overcome a $1/2$ BAE with a frequency of $23$ kHz is destabilized. In addition, the simulations with multiple EP simulations confirm that the main drivers of the EIC event are the helically trapped energetic particles generated by the perpendicular NBI, because the $1/1$ EIC are destabilized even if the tangential NBI injection intensity is small, but never if the perpendicular NBI injection intensity is below the $1/1$ EIC destabilization threshold.

Present study conclusions support the scenario proposed previously by other authors \cite{2,4}, identifying the EIC as resistive interchange modes further destabilized by the driving effect of the helically trapped EP. On the other hand, the destabilization of the $11/13$ EIC and the transition to the $1/1$ EIC when the EIC frequency chirp down during the bursting phase must be verified by dedicated experiments.

\ack

The authors would like to thanks Y. Todo, Y. Suzuki and A. Copper for fruitful discussions. We also thank the LHD technical staff for their contributions in the operation and maintenance of LHD.

\section*{Appendix}

\begin{figure}[h!]
\centering
\includegraphics[width=0.45\textwidth]{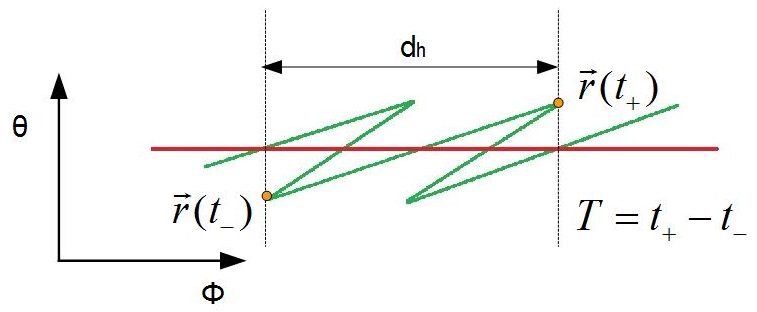}
\caption{Schematic view of the guiding center motion of the helically trapped EP that participate in the resonance. The red line indicates the magnetic field line and the green line the motion of the helically trapped EP. The yellow dots indicate the motion of the EP guiding center during one oscillation period.}\label{FIG:11}
\end{figure}

The effect of the helically trapped EP are introduced in the model through modification of the average drift velocity operator of the passing EP. Using the expressions 6.72, 6.77 and 6.83 of the ref \cite{74}, the trapped EP trajectory can be derived in the $\psi - \theta$ plane, described by the period of the guiding center motion in the helical ripple ($T$) and the adiabatic invariant $\Lambda$ defined as:

$$ \Lambda = 2 B_{\phi} \int^{\phi_{+}}_{\phi_{-}} \frac{m v_{||}}{B_{0}} d\phi = \frac{4 \psi^{`}}{a^{2} \sqrt{g}} \int^{\phi_{+}}_{\phi_{-}} \frac{m v_{||}}{B_{0}} d\phi $$ 
The averaged drift equations of the helically trapped EP can be written as:

$$ \frac{d\psi}{dt} = \omega_{b} \frac{4 \psi^{`}}{a^2} \frac{\partial}{\partial \theta} \left( \frac{1}{\sqrt{g}} \frac{m}{q} \int^{\phi_{+}}_{\phi_{-}} \frac{v_{||}}{B_{0}} d\phi \right) $$
$$ \frac{d\theta}{dt} = -\omega_{b} \frac{4 \psi^{`}}{a^2} \frac{\partial}{\partial \psi} \left( \frac{1}{\sqrt{g}} \frac{m}{q} \int^{\phi_{+}}_{\phi_{-}} \frac{v_{||}}{B_{0}} d\phi \right) $$
with $\omega_{b} = 1/2 \pi T$ the averaged bounce frequency of the guiding center. We can also define the square of the averaged bounce length of the guiding center ($d_{b}$), defined as:

$$ d_{b}^2 = \frac{m}{q} \int^{\phi_{+}}_{\phi_{-}} \frac{v_{||}}{B_{0}} d\phi = \frac{m}{q} \lim_{\Delta \phi \to\infty} \sum^{\phi_{+}}_{\phi_{-}} \frac{v_{||}}{B_{0}} d\phi $$
The experimental data indicates that the EP participating in the resonance have defined characteristics, thus we consider for simplicity the next assumption: the EP involved in the resonance have the same averaged guiding center bounce frequency and length. In effect, this implies that we are considering only particles at a single energy and pitch angle $\Delta = \mu B_{0} / 2$, with $\mu$ the magnetic moment. Consequently, as first order approximation, no radial or angular dependency of the averaged bounce length is included in the model, thus:

$$ \frac{d\psi}{dt} = \omega_{b} d_{b}^2 \frac{4 \psi^{`}}{a^2} \frac{\partial}{\partial \theta} \left( \frac{1}{\sqrt{g}} \right) $$
$$ \frac{d\theta}{dt} = -\omega_{b} d_{b}^2 \frac{4 \psi^{`}}{a^2} \frac{\partial}{\partial \psi} \left( \frac{1}{\sqrt{g}} \right) $$
Following such simplification, figure~\ref{FIG:11} shows a representation of the guiding center motion of the helically trapped EP that participate in the resonance.

If we define the averaged bounce velocity of the guiding center as $v_{b} = 2 \pi \omega_{b} d_{b}$, the averaged drift equations can be reformulated as:

\begin{eqnarray} 
v_{b} = \frac{\rho^{2}}{4 d_{b}} \left[ \int^{t_{+}}_{t_{-}} \frac{\partial}{\partial \theta} \left( \frac{1}{\sqrt{g}} \right) dt \right]^{-1} \nonumber\\
= - \frac{\theta}{4 d_{b}} \left[ \int^{t_{+}}_{t_{-}} \frac{\partial}{\partial \rho^{2}} \left( \frac{1}{\sqrt{g}} \right) dt \right]^{-1}
\end{eqnarray} 
where the Jacobian is an equilibrium variable thus it is not time dependent. Consequently:
$$\int^{t_{+}}_{t_{-}} \frac{\partial}{\partial \theta} \left( \frac{1}{\sqrt{g}} \right) dt = \frac{\partial}{\partial \theta} \left( \frac{1}{\sqrt{g}} \right)  \int^{t_{+}}_{t_{-}} dt = \frac{\partial}{\partial \theta} \left( \frac{1}{\sqrt{g}} \right) T$$
Using the expression of the averaged bounce velocity, we introduce a correction to the averaged drift velocity operator of the passing particles ($\Omega_{d,||}$):
$$ \frac{\epsilon^2 \pi \rho^2 \omega_{b}}{d_{b}} \left[ \frac{\partial}{\partial \theta} \left( \frac{1}{\sqrt{g}} \right) \right]^{-1} \Omega_{d,||} $$
The new averaged drift velocity operator includes information of the averaged bounce frequency and length of the guiding center of the helically trapped EP.

\hfill \break

\end{document}